\documentclass[aps,prb,amsmath,amssymb,reprint,superscriptaddress]{revtex4-2}

\usepackage{mathrsfs,amsfonts} 
\usepackage{bbm}
\usepackage[colorlinks,citecolor=blue,linkcolor=red,urlcolor=blue]{hyperref}
\usepackage{color}
\usepackage{graphicx}
\usepackage{subfigure}
\usepackage{verbatim}
\usepackage{fourier}
 \usepackage{soul}
 \setstcolor{red}

\begin{document}

\title{Near-field radiative heat transfer between a nanoparticle and a graphene grating}

\author{Minggang Luo}
\email[]{minggang.luo@umontpellier.fr}
\affiliation{Laboratoire Charles Coulomb (L2C), UMR 5221 CNRS-Universit\'e de Montpellier, F-34095 Montpellier, France}

\author{Youssef Jeyar}
\affiliation{Laboratoire Charles Coulomb (L2C), UMR 5221 CNRS-Universit\'e de Montpellier, F-34095 Montpellier, France}

\author{Brahim Guizal}
\affiliation{Laboratoire Charles Coulomb (L2C), UMR 5221 CNRS-Universit\'e de Montpellier, F-34095 Montpellier, France}


\author{Mauro Antezza}
\email[]{mauro.antezza@umontpellier.fr}
\affiliation{Laboratoire Charles Coulomb (L2C), UMR 5221 CNRS-Universit\'e de Montpellier, F-34095 Montpellier, France}
\affiliation{Institut Universitaire de France, 1 rue Descartes, Paris Cedex 05, F-75231, France}

\date{\today}

\begin{abstract}

We investigate the near-field radiative heat transfer between a normally and/or laterally shifted nanoparticle and a planar fused silica slab coated with a strip graphene grating. For this study we develop and use a scattering matrix approach derived from Fourier modal method augmented with local basis functions. 
We find that adding a graphene sheet coating on the slab can already enhance the heat flux by about 85\%. We show that by patterning the graphene sheet coating into a grating, the heat flux is further increased, and this happens thanks to the a topological transition of the plasmonic modes from circular to hyperbolic one, which allows for more energy transfer. The lateral shift affects the accessible range of high-$k$ modes and thus affects the heat flux, too. By moving the nanoparticle laterally above the graphene grating, we can obtain an optimal heat flux with strong chemical potential dependance above the strips. 
For a fixed graphene grating period ($D=1\mu$m) and not too large normal shift (separation $d<800$nm), two different types of lateral shift effects (e.g., enhancement and inhibition) on heat transfer have been observed. As the separation $d$ is further increased, the lateral shift effect becomes less important.  
%
%
We show that the lateral shift effect is sensitive to the geometric factor $d/D$. Two distinct asymptotic regimes are proposed: (1) the inhibition regime ($d/D<0.85$), where the lateral shift reduces the heat transfer, and (2) the neutral regime ($d/D \geq 0.85$) where the effect of the lateral shift is negligible. 
In general, we can say that the geometric factor $d/D \approx 0.85$ is a critical point for the lateral shift effect.
Our predictions can have relevant implications to the radiative heat transfer and energy management at the nano/micro scale.

\end{abstract}

\maketitle 

\section{introduction}
Recently, the study of near-field radiative heat transfer (NFRHT) has seen a surge in interest, motivated by both theoretical investigations and practical uses.
When the separation distance is comparable to or smaller than the thermal wavelength $\lambda_T=\hbar c/k_BT$,  near-field effects can cause the radiative heat flux to exceed the Planckian blackbody limit by several orders of magnitude \cite{Rytov1989,Polder1971,Loomis1994,Carminati1999}.
Theoretical investigations of NFRHT have extensively covered a variety of geometric configurations \cite{Shchegrov2000,Volokitin2001,Chapuis2008plate,Narayanaswamy2008,Chapuis2008,Manjavacas2012,Nikbakht2018,Messina2018,DongPrb2018,Zhang2019T,Luo2020prb_ensemble,Luo2023MTP,Luo2023IJHMT_diffusion,Biehs2011gratings,Kan2019prb,Zheng2022Materials_EMA}, some of which confirmed by pioneering experimental studies  \cite{Shen2009,Rousseau2009,Ottens2011,Song2015,Lim2015,Watjen2016,Ghashami2018,Yungui2018NC,DeSutter2019,Yungui2023PRAppl}.


Particularly, the excitation of high-order diffraction channels in gratings structures can significantly affect the fluctuational phenomena, such as the NFRHT \cite{Yang2017prl,Messina2017PRB,Hongliang_JHMT} and the Casimir interaction \cite{Messina2015pra_sphere_grating,Hobun,Casimir_Noto}. Additionally, the special electromagnetic conductivity features of graphene enables a wide modulation of the radiative heat transfer \cite{Svetovoy2012prb,Ognjen2012prb,Zheng2017,Volokitin2017Dey,Zhao2017prb,Shi2021am,LiuES2022,Lu2022small,GrafBilMA}, as well as other fluctuational phenomena, like the Casimir pressure in and out of thermal equilibrium \cite{Chahine2017prl,jeyar2024casimirlifshitz,Antezza_2024,Wang2024prb}.
%
%
It is interesting to explore whether the combination of grating geometry and the unique electromagnetic properties of graphene could induce new behaviors in NFRHT not observed with conventional materials.
Recent investigations on the NFRHT between two identical and perfectly aligned graphene gratings have shown that patterning the graphene sheets opens additional energy transfer channels \cite{Liu2015APL,Yang2020prAppl,Luo2023apl_NFRHT_grating}. Furthermore, the effect of twisting gratings on NFRHT has also been investigated \cite{He2020ol,He2020ijhmt,Luo2020apl}. It is worth noting that the effective medium theory, often used to study the NFRHT between substrate-supported graphene strips, treats the graphene grating as an homogeneous whole \cite{Zhou2022langmuir} and thus cannot account for the lateral shift effect. Therefore, a more accurate numerical method is needed to treat the NFRHT for structures with lateral shifts. More recently, a significant lateral shift induced modulation of the NFRHT between two identical misaligned graphene gratings has been reported \cite{Luo2024shifted}, where the Fourier modal method equipped with local basis functions (FMM-LBF) has been applied \cite{Guizal, Taiwan_LBF, PRE23,Luo2023Casimir_gg}.
%

On another note,  the NFRHT between nanoparticles and mere slabs has been studied in \cite{Joulain2014jqsrt_dipole,ChengQ2023ol,Fang2020prb_NP_slab,Fang2023IJHMT_NP_slab} but the situation between nanoparticles and gratings (two dissimilar structures) remains unexplored.   

\begin{figure} [htbp]
\centerline {\includegraphics[width=0.5\textwidth]{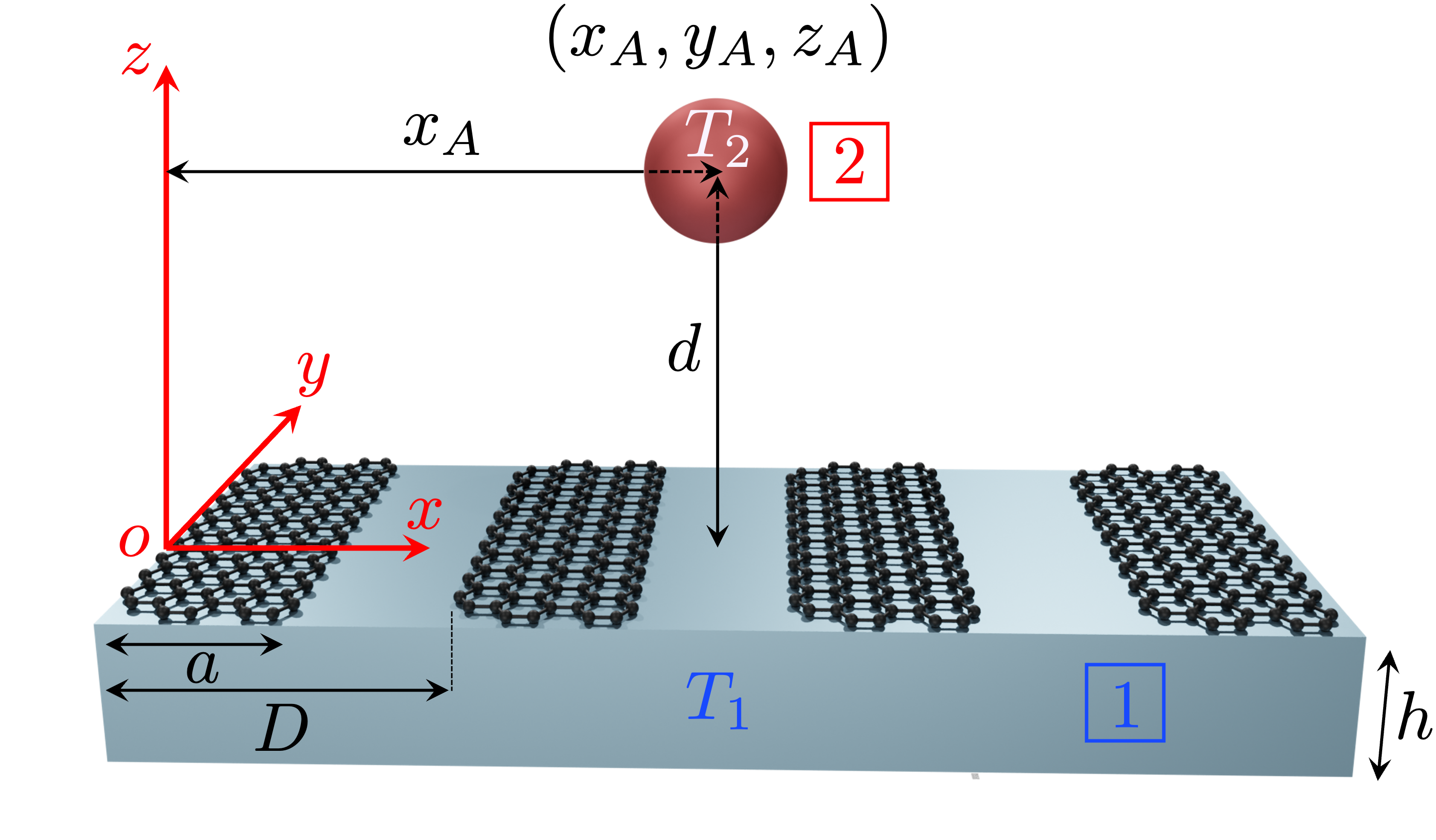}}
     \caption{Structure diagram of the supported graphene grating and a proximate small particle.}
   \label{fig:two_gratings_schematic}
\end{figure}

 In this work, we investigate the effect, of the lateral shift, on the NFRHT between a nanoparticle and a planar fused silica slab covered with a graphene grating (see Fig.~\ref{fig:two_gratings_schematic}).
%
 The paper is organized as follows: In Sec.~\ref{models}, we present the physical model for the nanoparticle and planar fused silica slab covered with graphene grating with relative normal and lateral shifts ($d$ and $x_A$), and the theoretical models using the scattering approach for radiative heat transfer. In Sec.~\ref{results_discussion}, we analyze the effects of various influencing factors (e.g., filling fraction, chemical potential, and grating period) on the lateral shift-mediated NFRHT and propose asymptotic regimes for the effect of normal and lateral shifts.

\section{Theoretical models}
\label{models}
The physical system under consideration is shown in Fig.~\ref{fig:two_gratings_schematic} where we consider the NFRHT between bodies 1 and 2 that are considered at fixed temperatures $T_1$ and $T_2$, respectively, while the environment is at the same temperature as body 1. Body 1 is a  planar fused silica (SiO$_2$) slab (with thickness $h$) covered with a graphene grating whose period is $D$ and where the width of one single graphene strip is $a$ (thus giving a filling fraction $f=a/D$). Body 2 is a small SiO$_2$ particle (with radius $R$). It  is located at position $\textbf{R}_A=(\textbf{r}_A,z_A) = (x_A,0,z_A)$ (we choose $y_A=0$ by virtue of the translation invariance along the strips in $y$-axis). The separation between bodies 1 and 2 is thus $d=z_A$.


The optical data of SiO$_2$ are taken from Ref.~\cite{Palik}, and the electromagnetic properties of graphene in body 1 are taken into account through its conductivity $\sigma_g$ which depends on temperature $T$ and on chemical potential $\mu$. It is the sum of an intraband and an interband contributions $\left(\sigma_g = \sigma_{\textnormal{intra}} +\sigma_{\textnormal{inter}}\right)$  given by \cite{Zhao2017prb,Falkovsky2007,Falkovsky2008,Awan_2016,rodriguezlopez2024graphene}:
%
\begin{equation}
\left\{
\begin{array}{rcl}
\begin{aligned}
&\sigma_{\textnormal{intra}} = \frac{i}{\omega + i/\tau} \frac{2 e^2 k_{\rm B} T}{\pi\hbar^2}\ln\left[ 2 \cosh\left(\dfrac{\mu}{2k_{\rm B}T}\right) \right]      ,
\\&\sigma_{\textnormal{inter}}= \frac{e^2}{4 \hbar} \left [ G\left(\frac{\hbar \omega}{2}\right) + i \dfrac{ 4\hbar \omega}{\pi}\int_0^{+\infty} \dfrac{G(\xi)-G\left(\hbar \omega /2\right)}{(\hbar \omega)^2 - 4\xi^2} {\rm d}\xi \right]	,
\end{aligned}
\end{array}
\right. 
\label{sigma_g}
\end{equation}
%
where  $e$  is the electron charge, $\tau$ the relaxation time (we use $\tau = 10^{-13}${s}), $G(\xi) = \sinh (\xi/k_B T) / [\cosh (\mu/k_{\rm B} T) + \cosh(\xi/k_{\rm B} T)]$, $\hbar$ is the reduced Planck constant, $\omega$ is the angular frequency, and $k_{\rm B}$ is the Boltzmann constant.\\

The net power flux $\varphi$ received by body 1 (energy per unit time) can be defined as \cite{Messina2014,Messina2011PRA}
%
\begin{equation}
\varphi= \sum_p \int \frac{{\rm d}^2 {\rm \textbf{k}}}{(2\pi)^2} \int_0^{+\infty} (\Theta(\omega,T_2)-\Theta(\omega,T_1)) \frac{{\rm d} \omega}{2\pi}<p,{\rm \textbf{k}}|\mathcal{O}|p,{\rm \textbf{k}}> ,
\label{HF}
\end{equation}
%
where  $p$ is the polarization index, $p=$ 1, 2 corresponding to TE (transverse electric) and TM (transverse magnetic) polarization modes respectively, $\Theta(\omega,T)=\hbar \omega/( e^{{\hbar \omega}/{k_{\rm B} T}}-1 )$ is the mean energy of the Planck oscillator, ${\rm \textbf{k}}=(k_x, k_y)$, $k=\sqrt{k_x^2+k_y^2}$, $k_x$ and $k_y$ being the wave vectors in the standard $(x, y, z)$ cartesian coordinates system. The transmission operator $ \mathcal{O}$ in the (TE, TM) basis is given by \cite{Messina2011PRA}:
\begin{widetext}
\begin{equation}
\mathcal{O}=   U^{(2,1)} \left[  f_{-1}(\mathcal{R}^{(2)-}) - \mathcal{T}^{(2)-} \mathcal{P}_{-1}^{({\rm pw})}  \mathcal{T}^{(2)-\dagger} \right] U^{(2,1)\dag } \left[  f_{1}(\mathcal{R}^{(1)+}) - \mathcal{T}^{(1)-\dagger} \mathcal{P}_{1}^{({\rm pw})}  \mathcal{T}^{(1)-} \right]  ,
\label{O_operator}
\end{equation}
\end{widetext}
where $U^{(2,1)}=\left(1-\mathcal{R}^{(2)-}\mathcal{R}^{(1)+}\right)^{-1}$, $\mathcal{R}^{(1)+}$ and $\mathcal{R}^{(2)-}$ ($\mathcal{T}^{(1)-}$ and $\mathcal{T}^{(2)-}$ ) are the reflection operators (transmission operators) of the grating and the nanoparticle in the (TE, TM) basis, $\dagger$ stands for the conjugation operation and the $\phi=\pm$ superscripts in the reflection and transmission coefficients correspond to the propagation direction with respect to the $z$ axis. $\left \langle p,{\textbf{k}}|\mathcal{P}_{\zeta}^{\rm (pw/ew)}|p',{\textbf{k}}' \right\rangle 
 =k_{z}^{\zeta}	\left \langle p,{\textbf{k}}|\mathcal{\prod}^{\rm (pw/ew)}|p',{\textbf{k}}' \right\rangle$, $k_{z}=\sqrt{k_0^2-{k}^2}$, $k_0=\omega/c$, $\mathcal{\prod}^{\rm (pw)}$ ($\mathcal{\prod}^{\rm (ew)}$) is the projector on the propagative (evanescent) sector, and the auxiliary function $f_{\zeta}(\mathcal{R})$ is given by \cite{Messina2014,Messina2011PRA}:
\begin{widetext}
\begin{equation}
f_{\zeta}(\mathcal{R})=\left\{
\begin{array}{rcl}
\mathcal{P}_{-1}^{(\rm pw)} - \mathcal{R} \mathcal{P}_{-1}^{(\rm pw)} \mathcal{R}^{\dagger} + \mathcal{R} \mathcal{P}_{-1}^{(\rm ew)}  -  \mathcal{P}_{-1}^{(\rm ew)} \mathcal{R}^{\dagger},   & & \zeta = -1, \\ \\
\mathcal{P}_{1}^{(\rm pw)} - \mathcal{R}^{\dagger} \mathcal{P}_{1}^{(\rm pw)} \mathcal{R} + \mathcal{R}^{\dagger} \mathcal{P}_{1}^{(\rm ew)}  -  \mathcal{P}_{1}^{(\rm ew)} \mathcal{R},   & & \zeta = 1.
\end{array}
\right.
\label{auxiliary_function}
\end{equation}
\end{widetext}

According to Ref.~\cite{Messina2017PRB}, the periodicity along the $x$ axis makes it natural to replace the mode variable $k_x$ with $k_{xn}=k_x+2\pi n/D$, $k_z$ becomes $k_{zn}=\sqrt{k_0^2- k_{xn}^2-k_y^2}$, and $k$ becomes $k_n=\sqrt{k_{xn}^2+k_y^2}$, where $n \in \mathbb{Z}$,  $k_{x}$ is in the first Brillouin zone $(-\frac{\pi}{D},\frac{\pi}{D})$, and $k_y \in \mathbb{R}$. Let's state, from now on, that during the numerical implementation, integer $n$ will be restricted to the interval $[-N,N]$, where $N$ is the so-called truncation order.

The calculation of the reflection and transmission operators $\mathcal{R}^{(1)+}$ and $\mathcal{T}^{(1)-}$ of body 1 (needed for our calculations) is obtained using the FMM-LBF and it is given in details in our recent work \cite{Luo2023apl_NFRHT_grating,Luo2023Casimir_gg,Luo2024shifted}.
Operators $\mathcal{R}^{(2)-}$ and ${\mathcal{T}}^{(2)-}$ of body 2 [small particle at position $(\textbf{r}_A,z_A)=(x_A,0,z_A)$] can be obtained within the dipole approximation (the separation $d$ is much greater than the particle radius \cite{Joulain2014jqsrt_dipole,Huth2010epjap,Mulet2001apl}) corresponding to the reference structure shown in Fig.~\ref{fig:two_gratings_schematic}.
%
 We have two types dipoles, \textit{i.e.}, an electric dipole with moment $\textbf{p}$ and a magnetic dipole with moment $\textbf{m}$. As reported in Ref.~\cite{Chapuis2014prb,Luo2019,DongPrb2017,Manjavacas2012}, the eddy-current Joule dissipation (of magnetic dipole origin) and the displacement current dissipation (of electric dipole origin) cooperate with each other to determine the heat dissipation in the particle. 
It has been shown, that in order to properly describe the radiative heat transfer between a metallic particle and  a metallic slab, it is necessary to use both  $\textbf{m}$ and $\textbf{p}$ \cite{Huth2010epjap,Chapuis2014prb}. However, in the case of a dielectric particle and dielectric slab, it is sufficient to use the dielectric moment $\textbf{p}$ \cite{Mulet2001apl,Joulain2014jqsrt_dipole,ChengQ2023ol}.
%
%

In this work, we aim to explore the radiative heat transfer between a dielectric SiO$_2$ particle and a dielectric SiO$_2$ slab covered with a graphene grating. We are safe to describe the dielectric SiO$_2$ particle as an electric dipole moment $\textbf{p}(\omega)=\varepsilon_0 \alpha(\omega) \textbf{E}(\textbf{R}_A,\omega)$, where $\alpha(\omega)$ is the polarizability of the small particle given by $\alpha(\omega)=4\pi R^3 [\varepsilon(\omega)-1]/ [\varepsilon(\omega)+2]$ in the Clausius Mossotti form \cite{Ben2011,Luo2022prb}, $\varepsilon_0$ is the vacuum permittivity, $\varepsilon(\omega)$ is the relative permittivity of the particle and $\textbf{E}(\textbf{R}_A,\omega)$ is the incident electric field at position $\textbf{R}_A$ at angular frequency $\omega$. Then, the reflection operator $\mathcal{R}^{(2)-}$ and the modified transmission operator $\tilde{\mathcal{T}}^{(2)-}$ can be written under the following form \cite{Messina2011PRA,Messina2014}
%
\begin{equation}
\left\{
\begin{array}{rcl}
\begin{aligned}
 \left \langle p,{\textbf{k}},n|\mathcal{R}^{(2)-}(\omega)|p',{\textbf{k}}',n' \right\rangle  = &\frac{i\omega^2\alpha(\omega)}{2c^2k_{zn}} \\ & \times \left[\hat{\textbf{e}}_p^-(\textbf{k},\omega,n) \cdot \hat{\textbf{e}}_{p'}^+(\textbf{k}',\omega,n') \right] \\ & \times  e^{i \left({k_{xn'}'}-{k_{xn}} \right) {x}_A}  e^{i \left( k_{zn}^{}+k_{z{n'}}' \right)z_A}	,
\\ \left \langle p,{\textbf{k}},n|\tilde{\mathcal{T}}^{(2)-}(\omega)|p',{\textbf{k}}',n' \right\rangle = & \frac{i\omega^2\alpha(\omega)}{2c^2k_{zn}} \\ & \times \left[\hat{\textbf{e}}_p^-(\textbf{k},\omega,n) \cdot \hat{\textbf{e}}_{p'}^-(\textbf{k}',\omega,n') \right] \\ &\times  e^{i \left({k_{xn'}'}-{k_{xn}} \right)  {x}_A}   e^{i\left(k_{zn}^{}-k_{zn'}'\right)z_A}	,
\end{aligned}
\end{array}
\right. 
\label{R2_T2_}
\end{equation}
%
where the polarization vectors $\hat{\textbf{e}}_{p}^{\phi}(\textbf{k},\omega,n) $ have the following expressions: 
\begin{equation}
\left\{
\begin{array}{rcl}
\begin{aligned}
 & \hat{\textbf{e}}_{\rm TE}^{\phi}(\textbf{k},\omega,n)=\frac{1}{k_n}(-k_y \hat{\textbf{x}}+k_{xn} \hat{\textbf{y}})   ,
\\& \hat{\textbf{e}}_{\rm TM}^{\phi}(\textbf{k},\omega,n)=\frac{1}{k_0} \left( -k_n \hat{\textbf{z}} + \phi k_{zn}  \frac{k_{xn}\hat{\textbf{x}}+k_{y}\hat{\textbf{y}}}{k_n}   \right)	.
\end{aligned}
\end{array}
\right. 
\label{e_TETM}
\end{equation}
In particular, in the limit of the absence of the small particle, $\mathcal{R}^{(2)-}=0$ and $\tilde{\mathcal{T}}^{(2)-} = 0$. According to the Ref.~\cite{Messina2011PRA,Messina2014}, we can write the transmission operator as the sum of the identity operator (describing the incoming field propagating unmodified) and of the modified transmission operator $\tilde{\mathcal{T}} $ accounting only for the scattered part of the field. Thus, the final transmission operator $\mathcal{T}^{(2)-}$ of the small particle becomes $\mathcal{T}^{(2)-}=\mathbb{1}+\tilde{\mathcal{T}}^{(2)-}$.

The effect of the relative lateral shift ($x_A$) and vertical separation ($d=z_A$) between bodies 1 and 2 on the radiative heat transfer is naturally included in the definitions of the reflection and transmission operators in Eq.~(\ref{R2_T2_}).

When the filling fraction $f=0$ or 1, body 1 reduces to the limit cases: bare slab ($f=0$) or graphene sheet covering the whole slab ($f=1$). In these two limit cases, the net radiative heat flux $\varphi$ between the bodies 1 and 2 can be divided into the propagating modes contribution and the evanescent modes contribution as \cite{Joulain2014jqsrt_dipole,Huth2010epjap,Fang2020prb_NP_slab,Fang2023IJHMT_NP_slab}
\begin{equation}
\varphi=\varphi_{\rm prop}+\varphi_{\rm evan}      ,
\label{HF_particle_plane_total}
\end{equation}
with
\begin{equation}
\begin{aligned}
\varphi_{\rm prop} = & \frac{2}{\pi}  \int_0^{+\infty} {\rm d} \omega (\Theta(\omega,T_2)-\Theta(\omega,T_1))k_0^2  \int_{k < k_0} \frac{{\rm d}^2 {\rm \textbf{k}}}{(2\pi)^2} \frac{1}{4k_{z0}} \\
&\times {\rm Im}(\alpha ) \left[(1-\vert r_{\rm s} \vert ^2 -\vert t_{\rm s} \vert ^2) + (1-\vert r_{\rm p} \vert ^2 -\vert t_{\rm p} \vert ^2) \right] ,
\end{aligned}
\label{HF_particle_plane_prop}
\end{equation}
and
\begin{equation}
\begin{aligned}
\varphi_{\rm evan} = & \frac{2}{\pi}  \int_0^{+\infty} {\rm d} \omega (\Theta(\omega,T_2)-\Theta(\omega,T_1))k_0^2  \int_{k > k_0} \frac{{\rm d}^2 {\rm \textbf{k}}}{(2\pi)^2} \frac{e^{-2\gamma z_A}}{2\gamma} \\
&\times  {\rm Im}(\alpha ) \left[{\rm Im} ( r_{\rm s} ) + \frac{2k^2-k_0^2}{k_0^2} {\rm Im} ( r_{\rm p} )  \right] ,
\end{aligned}
\label{HF_particle_plane_evan}
\end{equation}
where $\gamma=\sqrt{k^2-k_0^2}$ for the evanescent mode, $k_{z0}=\sqrt{k_0^2-k^2}$ for the propagative mode, $r_{\rm s}$ and $r_{\rm p}$ ($t_{\rm s}$ and $t_{\rm p}$) are the  ${\rm s}$- and ${\rm p}$-polarized Fresnel reflection (transmission) coefficients of body 1 [when there is a graphene sheet coating ({\it{i.e.}}, $f=1$) or without graphene ({\it{i.e.}}, $f=0$)] and are given in Refs. \cite{Shi2017Acs,prb_jeyar2023}.
%

\section{Results and discussion}
\label{results_discussion}
The separation $d$ between the particle and the slab is at least 5 times the particle's radius, which guarantees the validity of the dipole approximation for small particle \cite{Huth2010epjap,Joulain2014jqsrt_dipole}. The thickness of SiO$_2$ slab substrate $h=20$nm, $T_1=290$K, $T_2=310$K, SiO$_2$ particle radius $R=20$nm.

\subsection{Effect of graphene coating }
\begin{figure} [htbp]
\centerline {\includegraphics[width=0.5\textwidth]{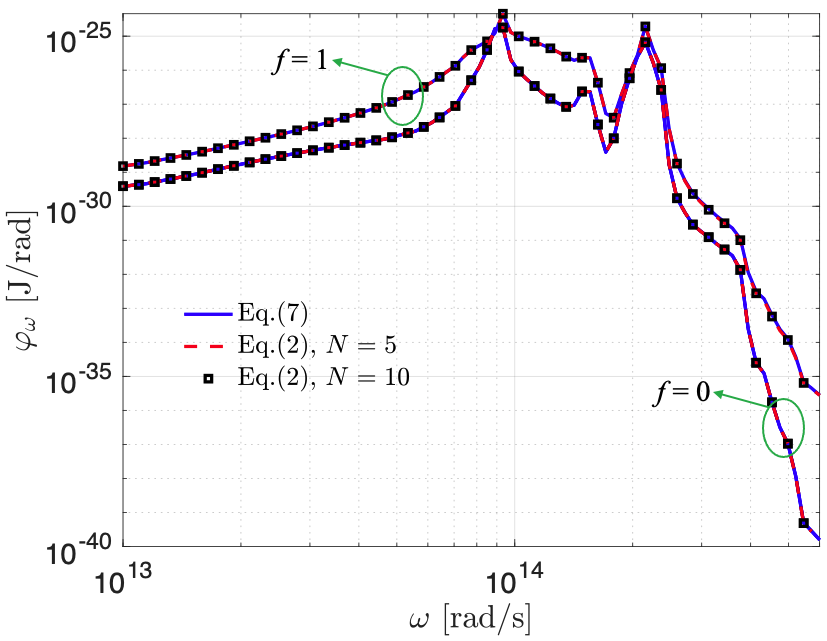}}
     \caption{Comparison between the radiative heat flux spectra obtained by two different methods: (1) by using Eq.(\ref{HF}) and fixing $f=0$ for a bare SiO$_2$ slab substrate (SiO$_2$ slab substrate with a full graphene sheet coating), and (2) by using Eq.(\ref{HF_particle_plane_total}). $d$=100nm.}
   \label{fig:verif_f_0_1}
\end{figure}

We start with the simple limit cases for the general coating, \textit{i.e.}, (1) Limit case 1: body 1 is a bare SiO$_2$ slab ($f=0$) with no coating, and (2) Limit case 2:  body 1 is a SiO$_2$ slab with a full graphene sheet coating ($f=1$).
%
%
The radiative heat flux spectra (HFS) $\varphi_{\omega}(\omega)$ of the two considered limit cases are shown in Fig.~\ref{fig:verif_f_0_1}.
We compare the HFS obtained by two different methods: (1) by using Eq.(\ref{HF}) and fixing $f=0$, and (2) by using Eq.(\ref{HF_particle_plane_total}) for the radiative heat flux between a small particle and a planar structure. The separation distance  $d$ is set to 100nm. The two different methods yield identical results, which means that the general Eq.(\ref{HF}) works accurately for the limit cases $f=0$ and $f=1$. In addition, when using the general grating-involved Eq.~(\ref{HF}) for these limit cases, two truncation orders have been used, $N=5$ and 10, to show that the convergence of HFS is already reached. That is, a small truncation order is sufficient to obtain an accurate result for the limit cases. However, due to the complex scattering details in the more general case ({\it i.e.}, $f \neq 0$ and $f \neq 1$), it is expected to have a different convergence behavior with respect to the truncation order.

\begin{figure} [htbp]
\centerline {\includegraphics[width=0.5\textwidth]{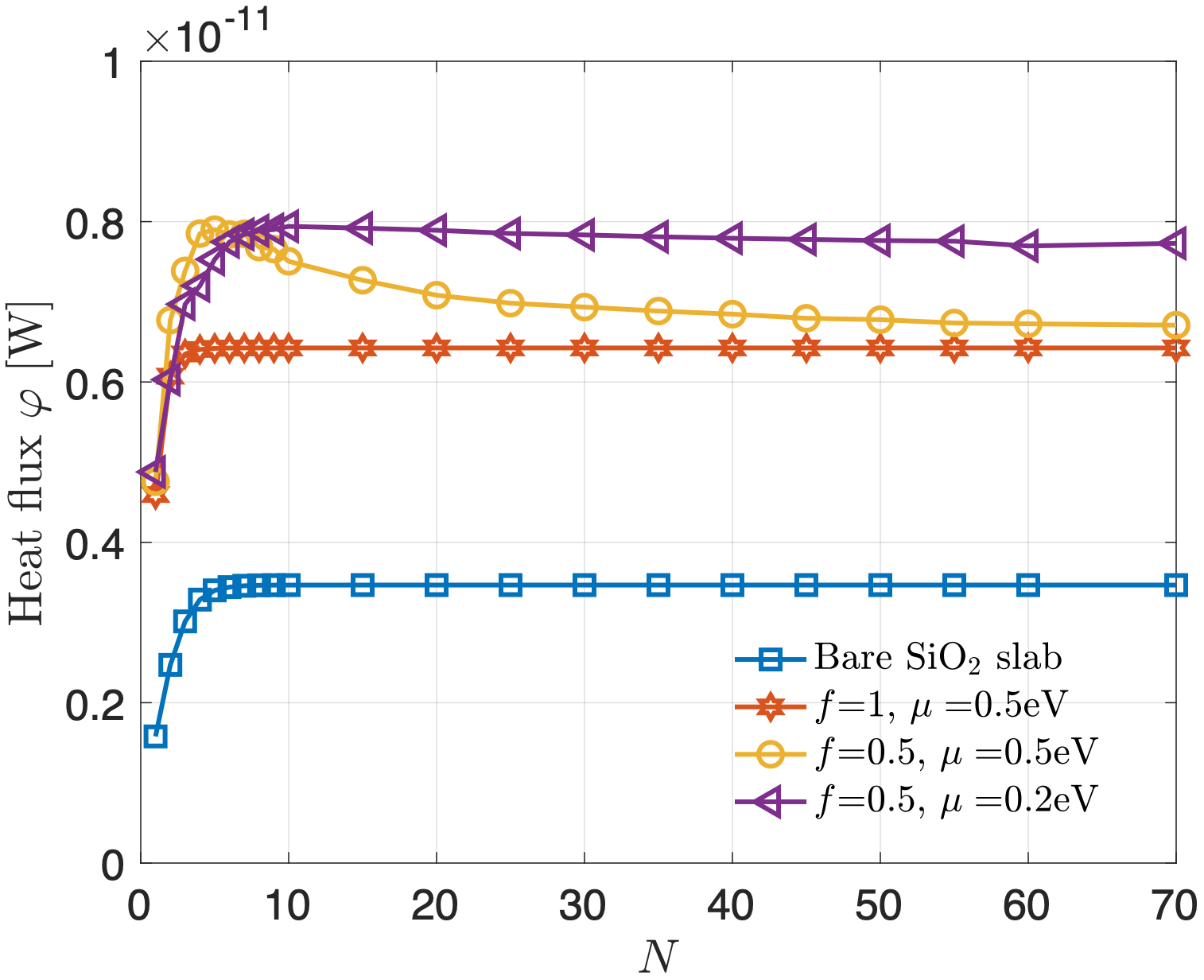}}
     \caption{Dependence of the radiative heat flux between 1 and 2 on the truncation order N. Four configurations are considered for body 1: (1) Bare SiO$_2$ slab ({\it{i.e.}}, $f=0$), (2) SiO$_2$ slab with a full graphene sheet coating ({\it{i.e.}},$f=1$), (3) SiO$_2$ slab covered with a graphene grating with filling fraction $f=0.5$ and chemical potential $\mu=0.5$eV, and  (4) The same configuration as (3) but with a different chemical potential $\mu=0.2$eV. Here: $d=100$nm, $x_A=0$nm and $D=1\mu$m.}
   \label{fig:trunc_depend}
\end{figure}

To see that, we consider the dependence of the heat flux $\varphi$ on the truncation order $N$ for the four cases shown in Fig.~\ref{fig:trunc_depend}: (1) Limit case 1, ({\it{i.e.}}, $f=0$), (2) Limit case 2,  ({\it{i.e.}}, $f=1$), (3) A general case with filling fraction $f=0.5$ and chemical potential $\mu=0.5$eV, and  (4) Another general case with filling fraction $f=0.5$ and chemical potential $\mu=0.2$eV. The separation distance is $d=100$nm, the lateral shift is $x_A=0$nm and the graphene grating period is $D=1\mu$m. We can see that for the two limit cases, the truncation order $N=10$ is sufficient to have a convergent result, which corresponds to the observations in Fig.~\ref{fig:verif_f_0_1}. However, for the general graphene grating coating cases, as expected, the convergence is slower. After many computational experiments, we found that the truncation order $N=40$ is sufficient to obtain convergent results. Therefore, in the following, we fix $N=40$ to guarantee convergent results (with a relative error less than 1\% on $\varphi$).

When a graphene sheet coating is applied to a bare SiO$_2$ slab, the heat flux between bodies 1 and 2 increases by about a factor of 1.85 compared to the bare slab case (See Fig.~\ref{fig:trunc_depend}). This enhancement is due to the contribution to heat transfer induced by the low-frequency graphene surface plasmon polaritons (SPPs) \cite{Ognjen2012prb,Svetovoy2012prb,Luo2023apl_NFRHT_grating} [See the significant increase of the HFS when coating the bare slab configuration ($f=0$) with a graphene sheet ($f=1$) in Fig.~\ref{fig:verif_f_0_1}]. This graphene coating induced enhancement of the heat flux between a small particle and planar structure is similar to that observed for two planar structures with graphene coating \cite{Svetovoy2012prb,Luo2023apl_NFRHT_grating,Luo2024shifted}. When patterning the graphene sheet into a grating ($f=0.5, \mu=0.5$eV, $d=100$nm, $D=1\mu$m), we can further enhance the heat flux by about 6\%. We also note that for the graphene grating coated configurations, varying $\mu$ from 0.5eV to 0.2eV brings another enhancement of the heat flux by about 16\%. 

\begin{figure} [htbp]
\centerline {\includegraphics[width=0.4\textwidth]{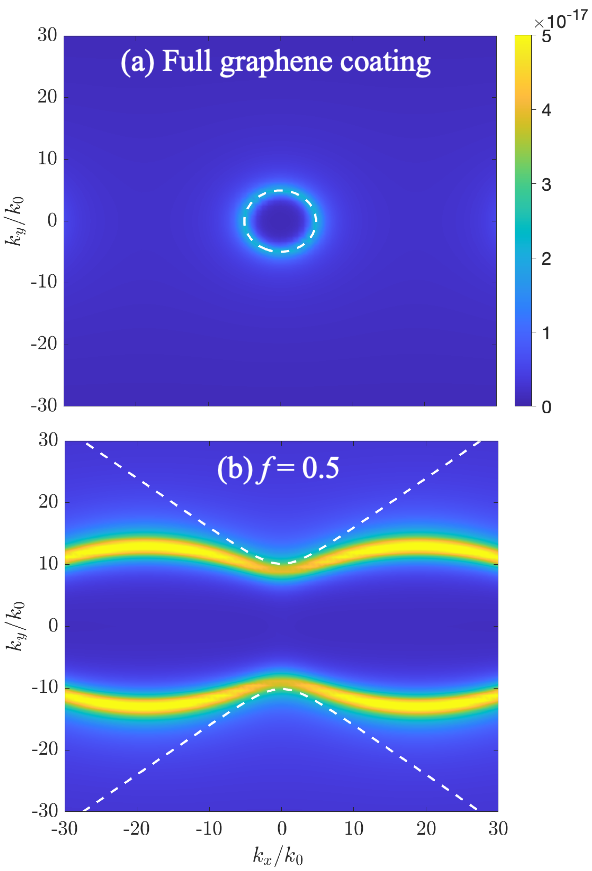}}
     \caption{Energy transmission coefficient for a SiO$_2$ particle and a SiO$_2$ substrate with two different coatings: (a) a graphene sheet coating, and (b) a graphene grating ($f=0.5$) coating. Here: $\mu=0.5$eV, $d$=100nm and $D=1\mu$m. The dotted curves represent the dispersion relations.}
   \label{fig:Energy_trans_f_0p5_to_f_1}
\end{figure}

To understand this patterning-induced enhancement, we introduce an energy transmission coefficient Tr($\mathcal{O}$) to describe the contributions of all polarizations to the heat flux in the ($k_x$,$k_y$) space. Tr($\mathcal{O}$) is the sum over the two polarizations of the photon tunneling probabilities, which is a counterpart of the one usually used to analyze the mechanisms behind the NFRHT between two gratings \cite{Greffet2012prb_trans_def,Liu2014apl_trans_def,Luo2023apl_NFRHT_grating,Luo2024shifted}. The transmission coefficient operator $\mathcal{O}$ is defined by Eq.~(\ref{O_operator}). In Fig.~\ref{fig:Energy_trans_f_0p5_to_f_1}, we show the energy transmission coefficient Tr($\mathcal{O}$) in the ($k_x/k_0$, $k_y/k_0$) plane at the angular frequency $\omega=5\times 10^{13}$rad/s (where the contribution of graphene to the NFRHT spectrum is important) for the system of a small particle and a slab with graphene coating. Two configurations are considered: (a) a SiO$_2$ particle and a SiO$_2$ substrate fully covered with a graphene sheet ($f=1$), and (b) the same structure but with a graphene grating coating (rather than a graphene sheet coating) with $f=0.5$, $\mu=0.5$eV, $d$=100nm and $D=1\mu$m. The dotted curves represent the dispersion relations obtained from the poles of the reflection coefficient \cite{Zhou2022langmuir,Zhou2022prm}. 

Similar to what happens in the the configuration of two identical graphene-involved planar structures studied in Refs. \cite{Liu2015APL,Luo2023apl_NFRHT_grating,Luo2024shifted}, the configuration of a nanoparticle and a graphene-involved finite planar slab, will,  upon graphene sheet patterning, lead to a topological transition for the surface plasmon polaritons from circular to hyperbolic one. For the graphene grating case corresponding to the hyperbolic modes, there are more accessible high-$k$ modes, while for the graphene sheet configuration (circular ones), the accessible wavevector region is relatively smaller. Consequently, the slab coated with a graphene grating gives rise to a greater transfer of energy compared to the slab coated with a graphene sheet.


\subsection{Effect of the filling fraction $f$ on  the NFRHT with lateral shift $\varphi(x_A)$}
\begin{figure} [htbp]
\centerline {\includegraphics[width=0.5\textwidth]{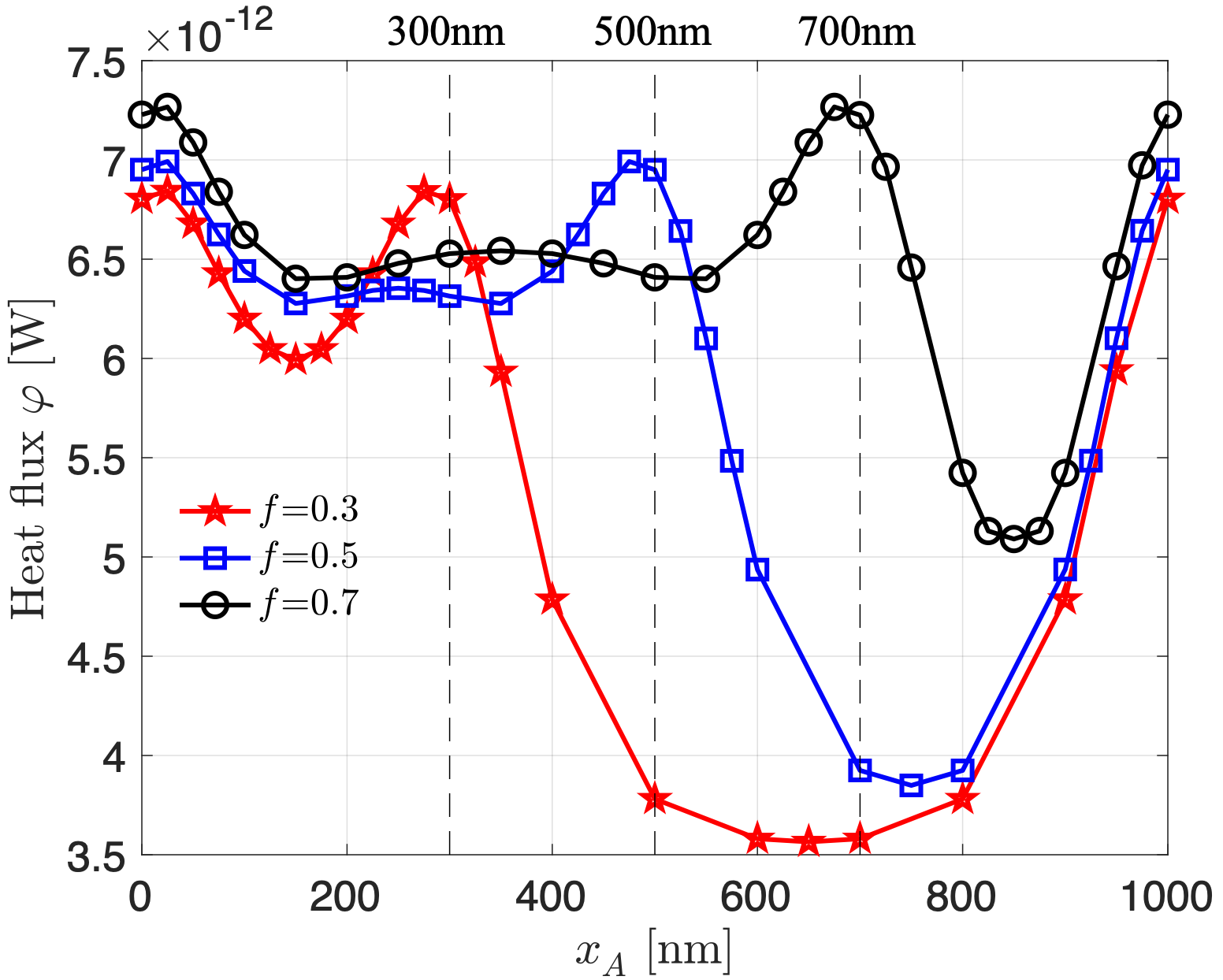}}
     \caption{Dependence of the radiative heat flux on the lateral shift $x_A$. Body 2 is an SiO$_2$ small particle. Three filling fractions are considered for the body 1: (1) $f=0.3$, (2) $f=0.5$, and (3) $f=0.7$. The width of one single graphene strip is shown as the dot line. $d$=100nm, $D=1\mu$m, $\mu=0.5$eV. }
   \label{fig:filling_frac_effect}
\end{figure}

In this section, we will discuss how the relative lateral shift $x_A$ affects the radiative heat transfer for different filling fractions $f$ of the graphene grating. The dependence of the heat flux $\varphi$ on the lateral shift $x_A$ is shown in Fig.~\ref{fig:filling_frac_effect} for a configuration with $D=1\mu$m, $d=100$nm and $\mu=0.5$eV. Three different filling fractions are considered, $f=0.3$, 0.5 and 0.7, that correspond to graphene strip widths $a=$300nm, 500nm and 700nm, respectively. The vertical dashed lines are added to indicate the corresponding graphene strip widths $a$.
%
%
%

The particle laterally shifts in one typical grating period ($0 < x_A < D$). The whole period $D$ can be divided into two regions: (1) the graphene strip region ($0 < x_A < a$) and (2) the bare SiO$_2$ slab region ($a < x_A < D$). The heat flux $\varphi$ varies when laterally shifting the particle in the whole period. In the two regions, the dependence of $\varphi$ on the lateral shift $x_A$ is very different. For a fixed grating period $D$ and a lateral shift $x_A$, changing the filling fraction $f$ (and thus the graphene strip width $a$) will change the scattering details, which results in a radically different heat flux $\varphi$ for the three configurations. {In the considered configurations, the variation range of heat flux is much larger in the bare SiO$_2$ slab region than that in the graphene strip region. In the graphene strip region, it is the graphene SPPs that contributes to the heat transfer, which does not vary too much when $x_A$ ranges from 0 (edge) to $a/2$ (strip center). However, in the bare SiO$_2$ slab region, when  $x_A$ ranges from $a$ (edge) to $(D+a)/2$ (slit center), the dominant contribution to heat transfer moves from the graphene SPPs to the surface phonon polaritons (SPhPs) supported by the substrate.}
%
%

\begin{figure*} [htbp]
\centerline {\includegraphics[width=1.\textwidth]{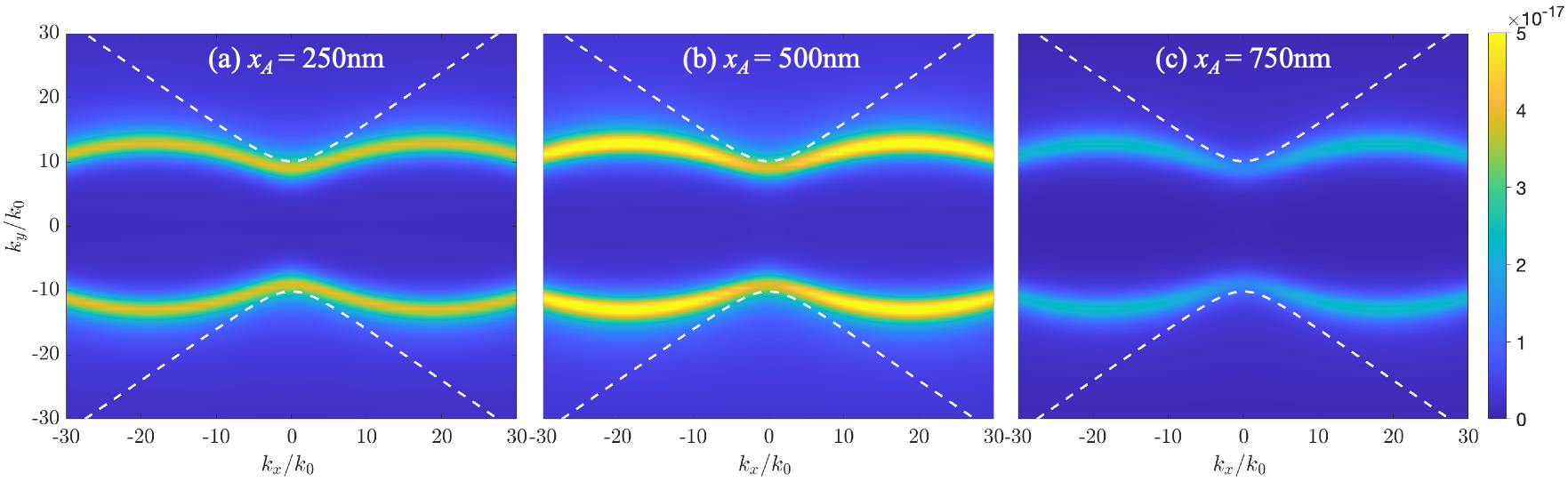}}
     \caption{Energy transmission coefficient for a SiO$_2$ particle and a graphene grating coated SiO$_2$ substrate  with three different lateral shifts: (a) $x_A=250$nm, (b) $x_A=500$nm, and (c) $x_A=750$nm. Here: $f=0.5$, $\mu=0.5$eV, $d=100$nm and $D=1\mu$m. The dotted curves represent the dispersion relations. }
   \label{fig:Energy_trans_f_0.5_different_xa}
\end{figure*}

To understand this lateral-shift-induced variation of the heat flux $\varphi$, we take the configuration with $f=0.5$, for example, and show the energy transmission coefficient Tr($\mathcal{O}$) in Fig.~\ref{fig:Energy_trans_f_0.5_different_xa} for three different lateral shifts, $x_A=250$nm, 500nm and 750nm and for $\omega=5\times 10^{13}$rad/s (where the contribution of graphene to the NFRHT spectrum is important). The other parameters are $\mu=0.5$eV, $d=100$nm and $D=1\mu$m. The dotted curves represent the dispersion relations obtained from the poles of the reflection coefficient \cite{Zhou2022langmuir,Zhou2022prm}. 
%
%

When shifting the particle from the center of the graphene strip region ($0 < x_A < a$) to the center of the bare SiO$_2$ slab region ($a < x_A < D$), the shape of the accessible mode region (the highlighted region in the figure) remains largely unchanged, but the value of the energy transmission coefficient for accessible modes changes significantly. The accessible range of high-$k$ modes for the configuration with $x_A=500$nm is the largest of the three configurations considered. The accessible range of high-$k$ modes for the $x_A=250$nm configuration (particle located at the center of region 1) is much larger than that for the $x_A=750$nm configuration (particle located at the center of region 2). Whether the lateral shift exists or not, the supported surface plasmon polaritons are alway the hyperbolic one. 
Whereas the patterning of graphene sheet into a grating brings a topology transition of the accessible range of high-$k$ modes, the lateral shift will affects these accessible range of high-$k$ modes for the configuration of a small particle and graphene-grating coated planar structure, which is similar to that observed for the configuration of two shifted identical graphene-gratings coated planar structures \cite{Luo2024shifted}.

\subsection{Effect of chemical potential $\mu$ on the NFRHT with lateral shift $\varphi(x_A)$}
\begin{figure} [htbp]
\centerline {\includegraphics[width=0.5\textwidth]{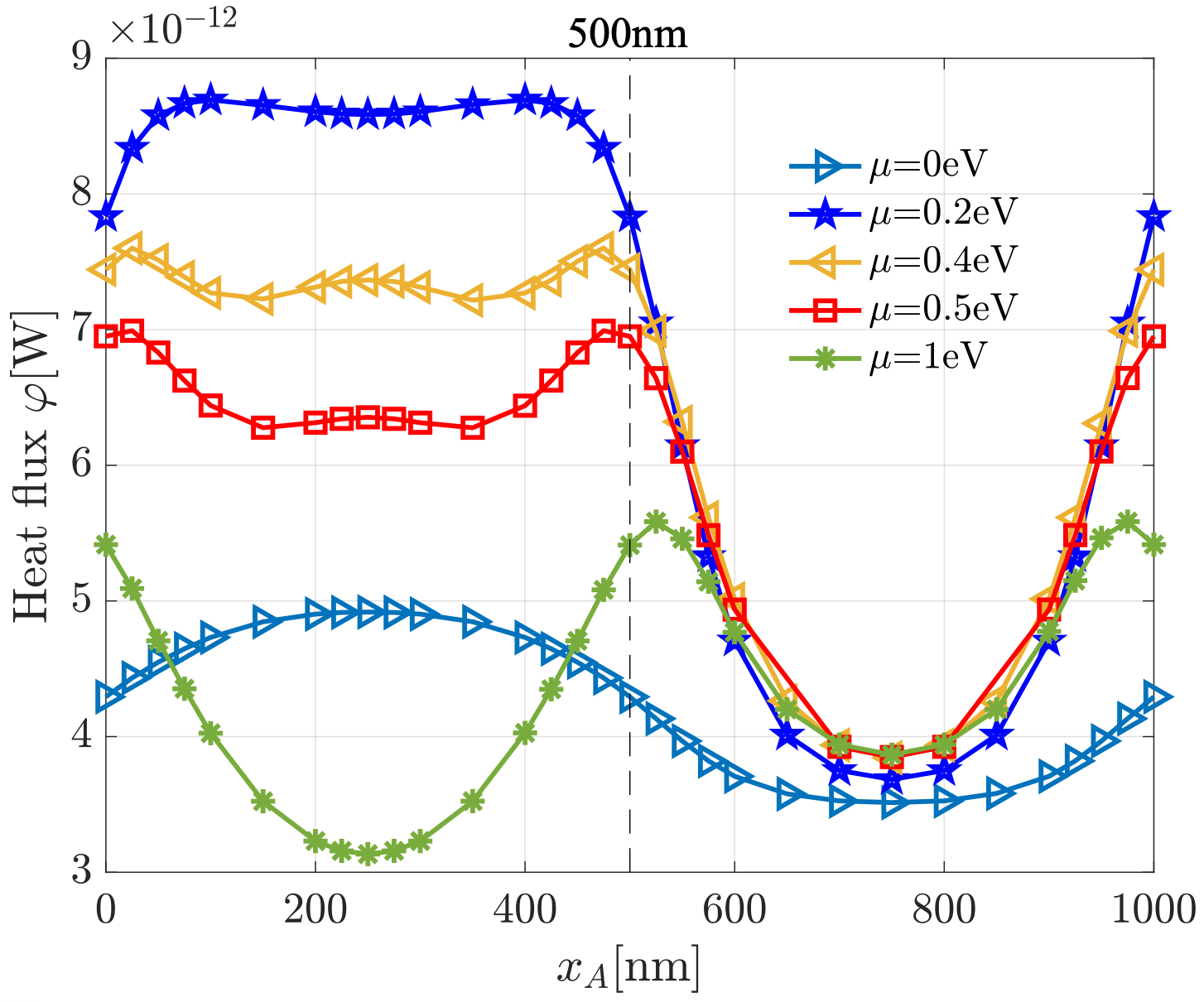}}
     \caption{Dependence of the radiative heat flux between on the lateral shift $x_A$. Here: $f=0.5$, $D=1\mu$m, $d=100$nm, five chemical potentials are considered for body 1, $\mu=0$eV, 0.2eV, 0.4eV, 0.5eV and 1.0eV. The dashed line marks the position corresponding to the strip width $a=fD=500$nm.}
   \label{fig:chemical_effect_a}
\end{figure}

In this section, we will discuss how the chemical potential $\mu$ gets involved in the lateral shift, induced modulation, of heat transfer. The dependence of the heat flux $\varphi$ on the lateral shift $x_A$ is shown in Fig.~\ref{fig:chemical_effect_a} for five configurations with different chemical potentials, $\mu=0$eV, 0.2eV, 0.4eV, 0.5eV and 1.0eV, respectively. Here, we took $D=1\mu$m and $d$=100nm. The dashed line marks the position corresponding to the strip width: $a=500$nm (in this case).
%
%

From Fig.~\ref{fig:chemical_effect_a}, we can see the dependence of the heat flux $\varphi(x_A)$ on the parameter that is the chemical potential $\mu$. Let's consider the two regions separated by the value $a$ and analyse the behaviour with respect to $\mu$ in each of them. For a fixed lateral shift $x_A$ in the graphene strip region ($0 < x_A < a$), when comparing the $\mu=0$eV configuration and $\mu=0.2$eV configuration, we see that the heat flux $\varphi$ increases rapidly with $\mu$. However, as $\mu$ is increased to 0.4eV, 0.5eV and 1.0eV, the heat flux $\varphi$ decreases gradually. We also note that $\varphi$ decreases to the valley and then rises back with increasing $x_A$ for the chemical potential $\mu \geq 0.4$eV. 
%
For a fixed lateral shift $x_A$ in the bare SiO$_2$ slab region ($a < x_A < D$), especially for the particle located at around the center of this region [$x_A \approx (D+a)/2$], the heat flux tends to be independent of the chemical potential. This is because in this region the particle is far away from the graphene coating part, where the SPhPs supported by the bare plate (certainly independent of the chemical potential $\mu$ of the graphene coating) dominate the heat transfer rather than the graphene SPPs. 

\begin{figure} [htbp]
\centerline {\includegraphics[width=0.5\textwidth]{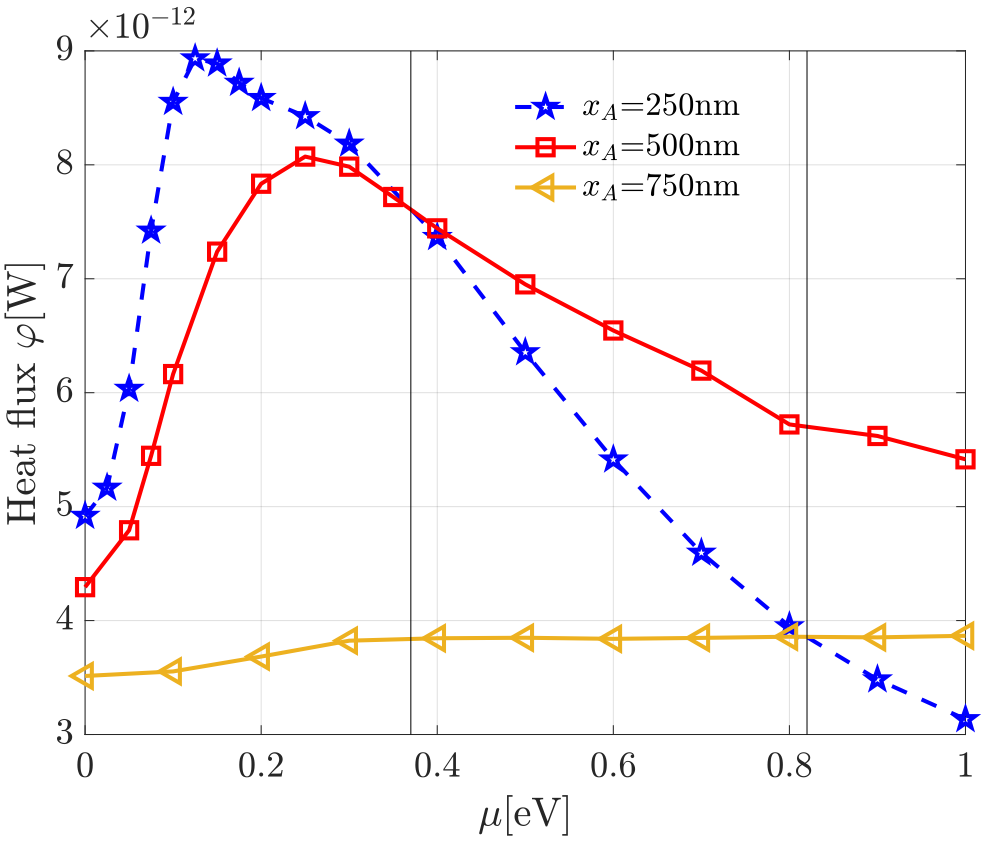}}
     \caption{Dependence of the radiative heat flux on the chemical potential $\mu$. Here: $f=0.5$, $d=100$nm, three different lateral shifts are considered, $x_A=250$nm, 500nm, and 750nm.}.  
   \label{fig:chemical_effect_b}
\end{figure}

Then, in order to see in more details the dynamics of $\varphi$ with respect to $\mu$, we show, in Fig.~\ref{fig:chemical_effect_b}, the dependence of the heat flux $\varphi$ on the chemical potential $\mu$ for three typical lateral shifts, $x_A=250$nm (center of the graphene coating region), 500nm (edge), and 750nm (center of the bare SiO$_2$ slab region).
For the $x_A=750$nm configuration, the maximal relative deviation $R_{\varphi} = (\varphi_{\rm max} - \varphi_{\rm min}) /  \varphi_{\rm min} \approx 10\%$, where $ \varphi_{\rm max} $ is at $\mu=1$eV and $ \varphi_{\rm min} $ is at $\mu=0$eV. The weak $\mu$-induced dependence of heat flux observed for the whole range of $\mu$ confirms that observed in Fig.~\ref{fig:chemical_effect_a}.
This relative deviation reaches $R_{\varphi} \approx$  88\% and 185\% for the $x_A=500$nm and $x_A=250$nm configurations, respectively. For  $x_A=500$nm, $ \varphi_{\rm max} $ is at $\mu=0.25$eV and $ \varphi_{\rm min} $ is at $\mu=0$eV, while for  $x_A=250$nm, $ \varphi_{\rm max} $ is at $\mu=0.125$eV and $ \varphi_{\rm min} $ is at $\mu=1$eV. When the particle is shifted in the graphene coating region, one can modulate the heat flux by simply tuning the chemical potential $\mu$. Furthermore, it is worth noting that the value of $\mu$ optimizing the heat flux varies with $x_A$. That is, a configuration of a different lateral particle shift will require a different $\mu$ to ensure an optimal heat flux.

%

In Fig.~\ref{fig:chemical_effect_b}, we can also see two crossing points whose positions are indicated by two vertical solid lines. The curves of the $x_A=250$nm and $x_A=500$nm configurations cross at the first crossing point at $\mu \approx 0.37$eV. When $\mu$ is fixed at the first crossing point, the heat flux changes only slightly when the particle is moved in the graphene coating region  ($0 < x_A < a$). 
This intersection point is also a critical point: (1) before this point the heat flux of the $x_A=250$nm configuration (particle located at the center of the graphene coating region) is larger than that of the $x_A=500$ (particle located at the edge), and (2) after this point the heat flux of the $x_A=250$nm configuration is smaller than that of the $x_A=500$nm.
The curves of configurations $x_A=250$nm and $x_A=750$nm cross at the second crossing point with $\mu \approx 0.82$eV. When moving the particle over the strip and the slit, the heat flux will decrease to the valley and then rise back. In general, the depths of the two valleys are different. It is worth noting that fixing $\mu$ at the second crossing point ($\mu \approx 0.82$eV) results in a nearly identical valleys depths for the heat flux curve in both regions.

\begin{figure} [htbp]
\centerline {\includegraphics[width=0.5\textwidth]{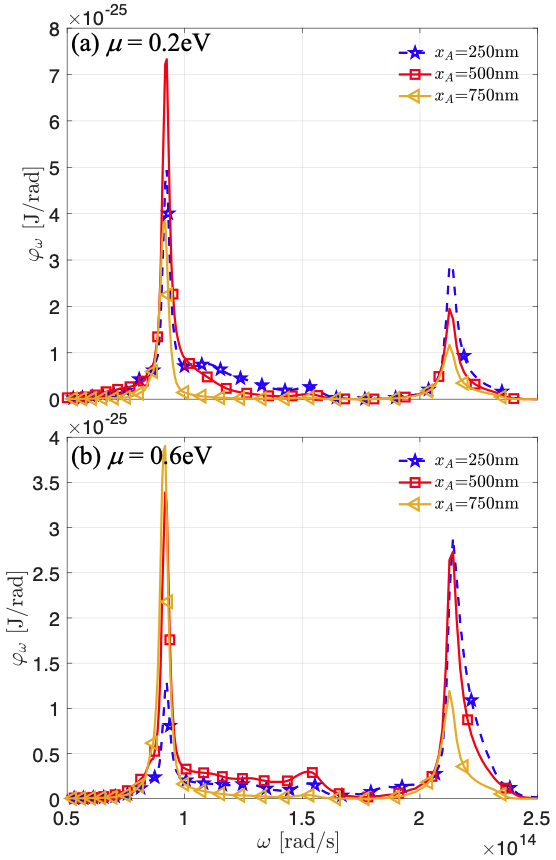}}
     \caption{Radiative heat flux spectra for two different chemical potentials: (a) $\mu=0.2$eV, and (b) $\mu=0.6$eV. Here:  $f=0.5$, $d=100$nm, three different lateral shifts are considered, $x_A=250$nm, 500nm, and 750nm.}
   \label{fig:chemical_effect_b_spectre}
\end{figure}

To understand the transition behavior around the critical point $\mu \approx 0.37$eV, we show the HFS in Fig.~\ref{fig:chemical_effect_b_spectre} for configurations of two different chemical potentials, $\mu=0.2$eV and 0.6eV, respectively. For the $\mu=0.2$eV configuration in Fig.~\ref{fig:chemical_effect_b_spectre} (a), between the two main peaks of the spectra, the curve of the $x_A=250$nm configuration is above that of the $x_A=500$nm configuration, which accounts for the higher heat flux for the $x_A=250$nm case than for the $x_A=500$nm case. However, the opposite happens for the $\mu= 0.5$eV configuration. In addition, for the whole range of angular frequency, the curve of the $x_A=750$nm configuration is below those of the other two configurations, which confirms the observations made in Fig.~\ref{fig:chemical_effect_b}.

%

\subsection{Asymptotic regimes for the normal and lateral shift effect on NFRHT}

\begin{figure} [htbp]
\centerline {\includegraphics[width=0.5\textwidth]{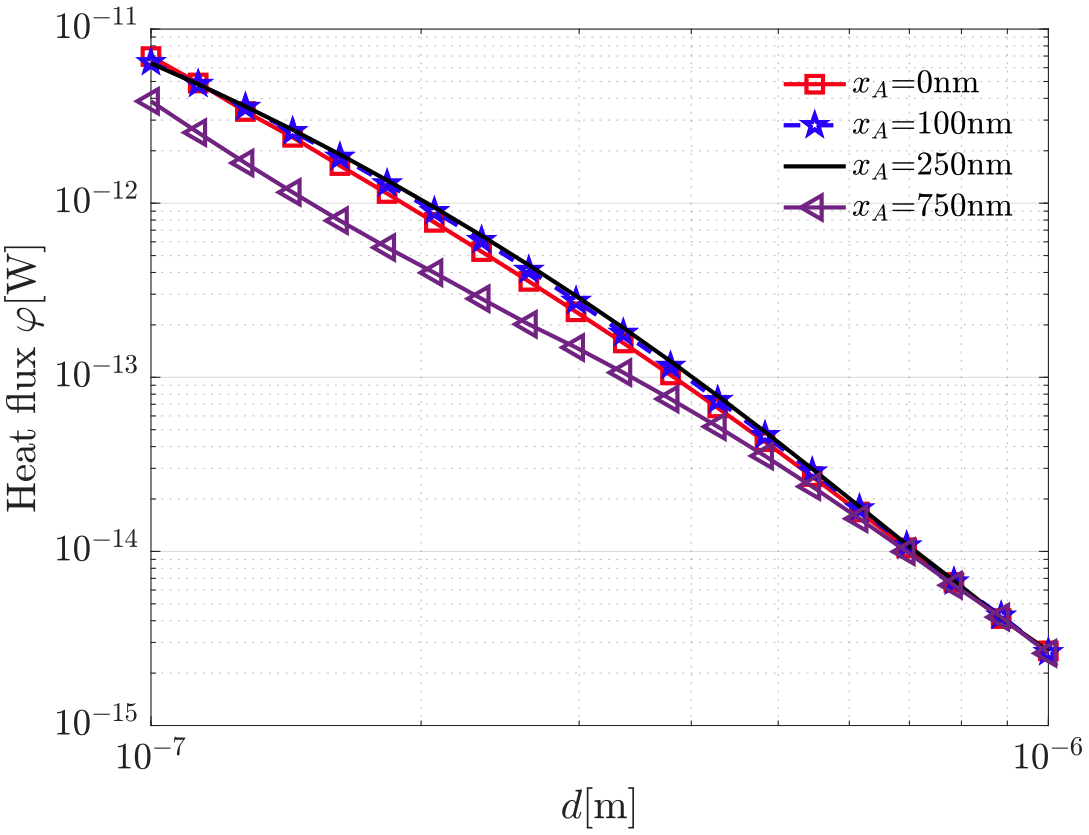}}
     \caption{Dependence of the radiative heat flux on the normal shift $d$. Four different lateral shifts are considered: $x_A=0$nm, 100nm, 250nm and 750nm. $D=1\mu$m, $\mu=0.5$eV, $f=0.5$.}
   \label{fig:separation_dependence}
\end{figure}

In this section, we consider a nanoparticle and an SiO$_2$ slab coated with a graphene grating and discuss the dependence of the NFRHT between on the normal shift of the particle, and that for different lateral shifts. We will try to propose an asymptotic regime map to quickly tell the existence of the lateral shift effect or not, which is a natural extension of our previous work on the asymptotic regime map for the two identical graphene gratings configuration \cite{Luo2024shifted}. To start, the dependence of the heat flux on the normal shift (separation) $d$ is shown in Fig.~\ref{fig:separation_dependence} for a few separate and different lateral shifts, $x_A=0$nm, 100nm, 250nm, and 750nm, where $D=1\mu$m, $\mu=0.5$eV, $f=0.5$, and the normal shift (separation) $d \in (100{\rm nm} ,1000{\rm nm})$.

%

For large normal separations ($d > 800$nm), configurations with four different lateral shifts show a quasi-identical heat flux. That is, when $d>800$nm, the nanoparticle sees the graphene grating as an effective whole rather than a grating with its details, and thus the lateral shift effect on the heat transfer becomes less important. For short normal separations ($d < 800$nm), the dependence of the heat flux on $d$ is different within each configuration of the considered four different lateral shifts $x_A$.  In a wide range of short normal separations, compared with the result of the configuration without a lateral shift (red square curve), we can see two distinct kinds of lateral shift effects on the heat flux: (1) the favorable one (black solid curve), corresponding to the configuration with $x_A=250$nm (center of graphene strip), and (2) the inhibitive one (purple left triangle curve), corresponding to the configuration with $x_A=750$nm (center of bare SiO$_2$ slab region). It is worth noting that we still can observe an exception for the very short cases (approximately $d=100$nm), i.e., no favorable lateral shift effect is present in the figure.
According to Fig.~\ref{fig:filling_frac_effect}, when shifting the nanoparticle around the center of the graphene strip region in the configuration of a fixed normal shift ($d=100$nm), the heat flux only shows a slight deviation. This observation is shown to be more general for a wider range of normal shifts in Fig.~\ref{fig:separation_dependence}, where we find that the dependence of the heat flux $\varphi$ on $d$ is similar between the $x_A=100$nm and $x_A=250$nm configurations.

\begin{figure} [htbp]
\centerline {\includegraphics[width=0.5\textwidth]{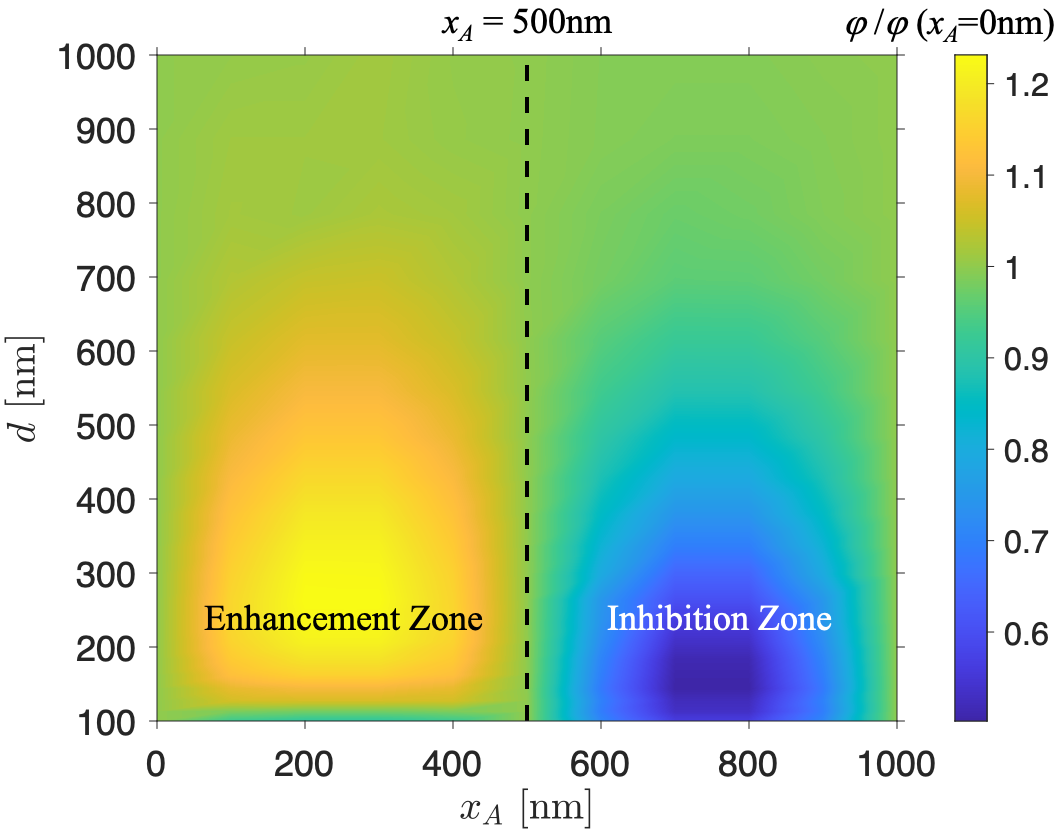}}
     \caption{Dependence of $\varphi/\varphi(x_A=0{\rm nm})$ on the normal shift $d$ and lateral shift $x_A$. $D= 1\mu$m, $\mu=0.5$eV, $f=0.5$. $x_A=a=500$nm is shown as the dashed line. $a$ is single graphene strip width.}
   \label{fig:separation_shift_contour}
\end{figure}

To clearly show the effect, on the heat flux, of shifting the nanoparticle in both $x$ and $z$ directions, we show, in Fig.~\ref{fig:separation_shift_contour}, the ratio $\varphi/\varphi(x_A=0{\rm nm})$ in the plane ($x_A,d$) for 
$D= 1\mu$m, $\mu=0.5$eV and $f=0.5$. The dashed line $x_A=500$nm is added for reference. Along the $x$-axis, the nanoparticle is shifted over one full grating period, i.e., $0<x_A<D$ while along the $z$-axis, it is normally shifted in the range $100{\rm nm}(=0.1D)<d<1000{\rm nm}(=D)$.
 According to the value of the  ratio $\varphi/\varphi(x_A=0{\rm nm})$ in this figure, we can tell out two distinct zones: (1) An enhancement Zone [the ratio $\varphi/\varphi(x_A=0{\rm nm})>1$, the lateral shift improves the heat transfer], and (2) An inhibition Zone [the ratio $\varphi/\varphi(x_A=0{\rm nm})<1$, the lateral shift hinders the heat transfer]. 
For a sufficiently large normal shift of the nanoparticle ($d>800$nm), these two zones are no longer apparent. 
 It is also important to remark that shifting the nanoparticle along the $y$-axis does not affect the heat flux between the two bodies, according to the invariance of body 1 along this axis. 
%
%

\begin{figure} [htbp]
\centerline {\includegraphics[width=0.5\textwidth]{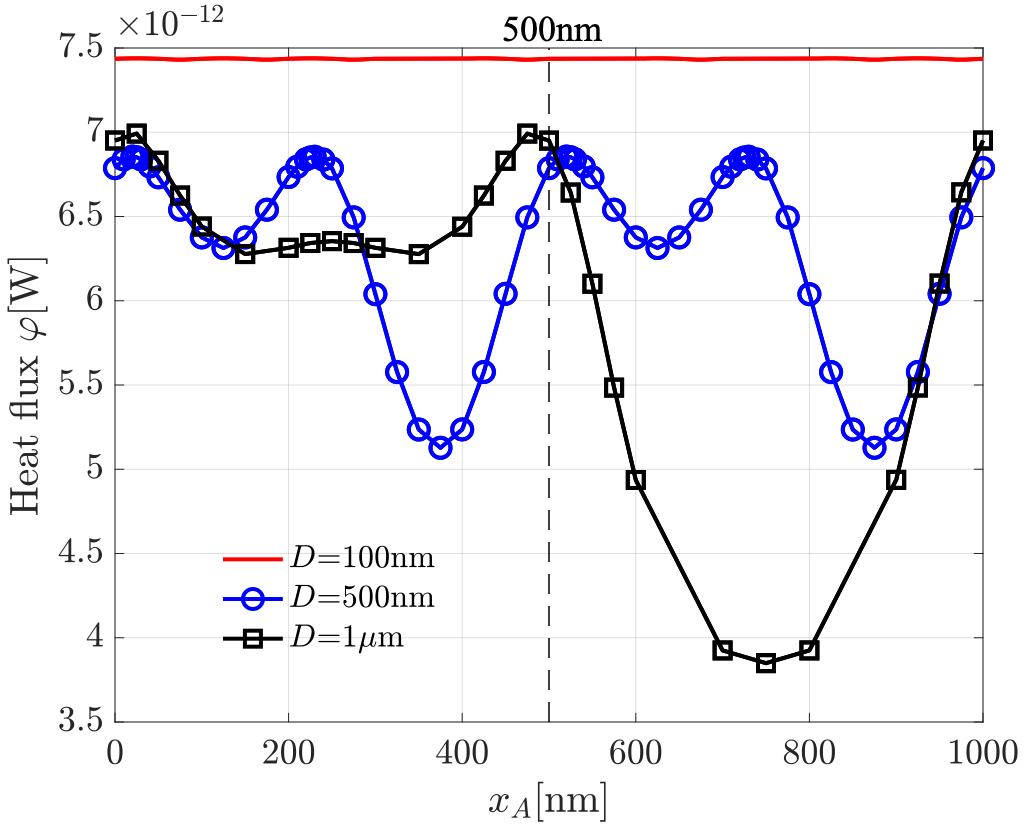}}
     \caption{Dependence of the radiative heat flux on $x_A$. Three different grating periods are considered: $D=100$nm, 500nm, and 1$\mu$m, respectively, with $d=100$nm, $f=0.5$, and $\mu=0.5$eV. The width of one single graphene strip of the configuration with $D=1\mu$m is shown as the dashed line.  }
   \label{fig:period_effect}
\end{figure}

We then show, in Fig.~\ref{fig:period_effect}, the dependence of the radiative heat flux on $x_A$ for three different grating periods: $D=100$nm, 500nm, and 1$\mu$m, respectively, with $f=0.5$ and $\mu=0.5$eV. The dashed line marking the position corresponding to $a=500$nm (for $D=1\mu$m) is added for reference.
%


For a normal shift fixed at $d=100$nm, as the period $D$ increases from 100nm to 1$\mu$m, the dependence of the heat flux $\varphi$ on $x_A$ changes significantly. For the $D=100$nm configuration, the heat flux $\varphi$ remains nearly constant as the nanoparticle is shifted laterally. However, for the other two configurations ($D=500$nm and $D=1\mu$m), the lateral shift significantly affects the heat flux. That is, for a fixed normal shift (separation) $d$, when the period of graphene grating $D$ is gradually decreased, the nanoparticle tends to see the grating as an effective whole whose structure invariance occurs in the plane (not only along the $y$-axis but also along the $x$-axis). 

A natural question is the existence of a critical grating period $D$, for the configurations with a fixed normal $d$, where the lateral shift effect disappears. Furthermore, is it possible to propose a regime map of the lateral shift effect for the heat transfer between the nanoparticle and graphene grating coated structure, similar to that of the two identical graphene grating coated slabs reported in our previous work \cite{Luo2024shifted}? 

\begin{figure} [htbp]
\centerline {\includegraphics[width=0.5\textwidth]{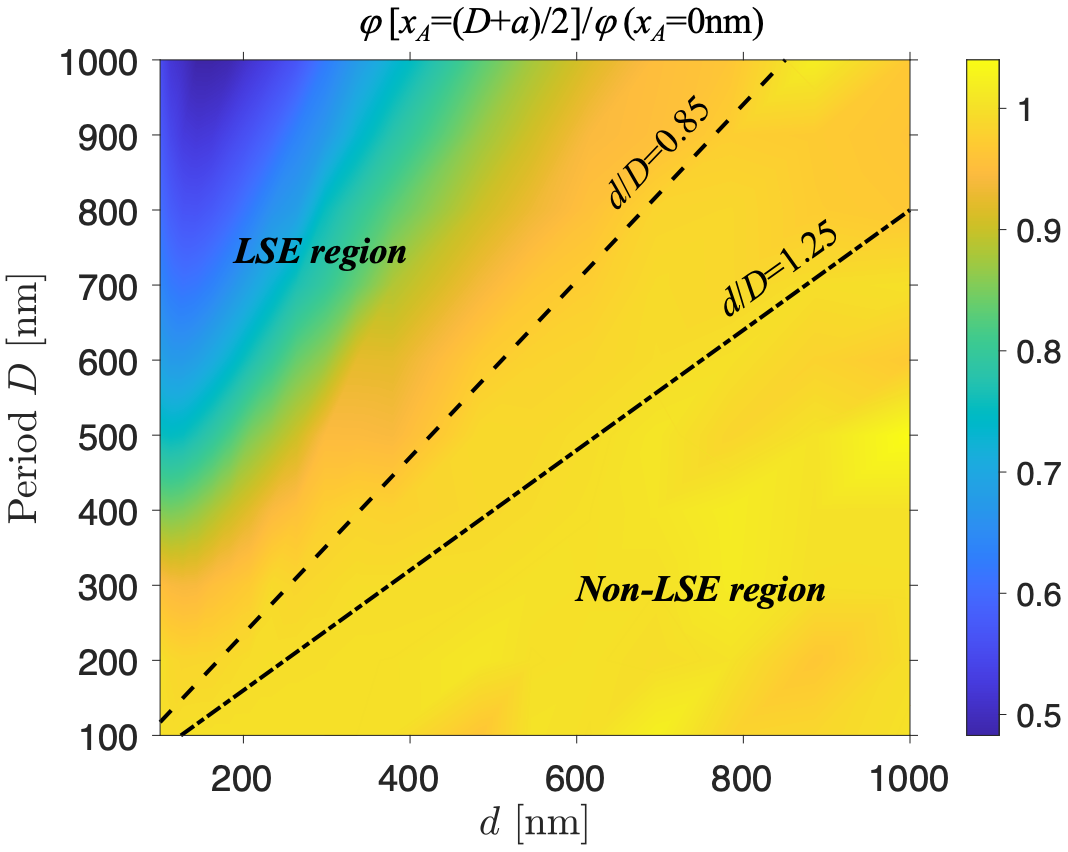}}
     \caption{Dependence of the ratio $ \varphi \left[ x_A = (D+a)/2 \right] / \varphi (x_A=0{\rm nm})$ on the normal shift $d$ and period $D$. Here: $f=0.5$ and $\mu=0.5$eV. The lines (geometric factor$d/D=0.85$ and 1.25, respectively) are added for reference.}
   \label{fig:separation_period_contour}
\end{figure}

To answer these questions, following the method used in \cite{Luo2024shifted}, we use the geometric factor $d/D$ to unveil the lateral shift effect regime map.
In figure  \ref{fig:separation_period_contour}, we show the ratio $\varphi \left[ x_A = (D+a)/2 \right] / \varphi (x_A=0{\rm nm}) $ in the plane ($d,D$) with $f=0.5$ and $\mu=0.5$eV. The lines [geometric factor $d/D=0.85$(dashed line) and $d/D=1.25$ (dash-dotted line), respectively] are added for reference.
According to the value of the  ratio $\varphi \left[ x_A = (D+a)/2 \right] / \varphi (x_A=0{\rm nm}) $ in the figure, we can tell out two distinct regimes: (1) the lateral shift effect regime [\textit{LSE region}, $d/D<0.85$, the ratio $\varphi \left[ x_A = (D+a)/2 \right] / \varphi (x_A=0{\rm nm})  < 1$, the lateral shift is against the heat transfer], and (2) the negligible lateral shift effect regime [\textit{non-LSE region}, $d/D \geq 0.85$, the ratio $\varphi \left[ x_A = (D+a)/2 \right] / \varphi(x_A=0{\rm nm}) \approx 1$]. 
In general, we can say that the geometric factor $d/D \approx 0.85$ is a critical point for the lateral shift effect.

%

\section{conclusion}

We studied the effect of normal and lateral shifts on NFRHT between a nanoparticle and a finite-thickness planar fused silica slab coated with a graphene grating by using the FMM-LBF.  
Graphene sheet coating on a slab can enhance the heat flux by about 85\%. By patterning the graphene sheet coating into a grating, the heat flux will increase further by about 6\% (e.g., $f = 0.5$, $\mu = 0.5$eV, $d = 100$nm, $D = 1\mu$m), which is due to the a topological transition for the accessible modes from circular to hyperbolic one, allowing for more energy transfer. 
When laterally shifting the nanoparticle over one period, according to the effect of lateral shift on heat flux, we observed two regions: (1) graphene strip region (14\% reduction of heat flux, see $f=0.3$ configuration in Fig.~\ref{fig:filling_frac_effect}) and (2) bare SiO$_2$ slab region (50\% reduction of heat flux). The heat flux variation induced by the lateral shift in these two regions is different in each case. The lateral shift significantly affects the value of the energy transmission coefficients, but does not affect the shape of the accessible region for the high-$k$ modes and thus affects the heat flux, unlike the patterning of graphene sheet into a grating which brings a topology transition of the accessible range of high-$k$ modes.
%
In general, the chemical potential affects significantly the heat flux, however, the chemical potential dependence of the heat flux is different for each lateral shift. When laterally shifting the nanoparticle in the former region, we can get an optimal (peak) heat flux by tuning the chemical potential, whose peak position changes with $x_A$. When laterally shifting the nanoparticle in the second region, the chemical dependence becomes weak and the heat flux tends to be constant.

For a fixed graphene grating period ($D=1\mu$m) and not too large normal shift (separation $d<800$nm), the lateral shift effect on heat transfer is of two types, enhancement and inhibition. When increasing further the separation $d$, the lateral shift effect becomes less important. We also show that for a fixed normal shift ($d=100$nm), the lateral shift dependence of the heat flux is different when changing the grating period. For $D=100$nm ($d/D=1$), the heat flux remains nearly constant when laterally shifting the nanoparticle. As the period ($D$) increases ($d/D$ decreasing), the lateral shift effect becomes more and more important.  
We find that the lateral shift effect is sensitive to the geometric factor $d/D$ (see Fig.~\ref{fig:separation_period_contour} for the regime map of the lateral shift effect). 
According to the value of the  ratio $ \varphi \left[ x_A = (D+a)/2 \right] / \varphi (x_A=0{\rm nm})$ in the figure, we can distinguish two regimes: (1) the lateral shift effect regime [ $d/D<0.85$, the ratio $ \varphi \left[ x_A = (D+a)/2 \right] / \varphi (x_A=0{\rm nm}) < 1$, the lateral shift is against the heat transfer], and (2) the negligible lateral shift effect regime [$d/D \geq 0.85$, the ratio $\varphi \left[ x_A = (D+a)/2 \right] / \varphi(x_A=0{\rm nm}) \approx 1$]. 
In general, we can say that the geometric factor $d/D \approx 0.85$ is a critical point for the lateral shift effect.

\begin{acknowledgments}
This work was supported by a grant "CAT" (No. A-HKUST604/20) from the ANR/RGC Joint Research Scheme sponsored by the French National Research Agency (ANR) and the Research Grants Council (RGC) of the Hong Kong Special Administrative Region, China.
\end{acknowledgments}

\bibliography{HT}

\end{document}